\documentclass[%
 aip,
 amsmath,amssymb,
preprint,%
]{revtex4-1}

\usepackage{graphicx}
\usepackage{dcolumn}
\usepackage{bm}

\usepackage[utf8]{inputenc}
\usepackage[T1]{fontenc}
\usepackage{mathptmx}
\usepackage{etoolbox}
\usepackage{color}

\makeatletter
\def\@email#1#2{%
 \endgroup
 \patchcmd{\titleblock@produce}
  {\frontmatter@RRAPformat}
  {\frontmatter@RRAPformat{\produce@RRAP{*#1\href{mailto:#2}{#2}}}\frontmatter@RRAPformat}
  {}{}
}%
\makeatother
\begin{document}


\title{Low-temperature electron dephasing rates indicate magnetic disorder in superconducting TiN films}

\author{A.\,I.\,Lomakin}
\affiliation{National University of Science and Technology MISIS, Moscow, Russia}
\author{E.\,M.\,Baeva}
\affiliation{Moscow Pedagogical State University, Moscow, Russia}
\affiliation{HSE University, Moscow, Russia}
\author{N.\,A.\,Titova}
\author{A.\,V.\,Semenov}
\affiliation{Moscow Pedagogical State University, Moscow, Russia}
\author{A.\,V.\,Lubenchenko}
\affiliation{National Research University MPEI, Moscow, Russia}
\author{M.\,A.\,Kirsanova}
\affiliation{Advanced Imaging Core Facility, Skolkovo Institute of Science and Technology, Moscow, Russia}
\author{S.\,A.\,Evlashin}
\affiliation{Center for Materials Technologies, Skolkovo Institute of Science and Technology, Moscow, Russia}
\author{S. Saha}
\affiliation{Argonne National Laboratory, Lemont, USA }
\author{S.\,Bogdanov}
\affiliation{Department of Electrical and Computer Engineering, University of Illinois at Urbana-Champaign, Urbana, USA}
\affiliation{Holonyak Micro and Nanotechnology Lab, University of Illinois at Urbana-Champaign, Urbana, USA}
\affiliation{Illinois Quantum Information Science and Technology Center, University of Illinois at Urbana-Champaign, Urbana, USA}
\author{A.\,I.\,Kolbatova*}
\email[Corresponding author: ]{aikolbatova@gmail.com}
\affiliation{Moscow Pedagogical State University, Moscow, Russia}
\author{ G.\,N.\,Goltsman}
\affiliation{Moscow Pedagogical State University, Moscow, Russia}
\affiliation{HSE University, Moscow, Russia}

\date{\today}

\begin{abstract}
We investigate electron transport and phase-breaking processes in thin titanium nitride (TiN) films of epitaxial quality. Previous studies show that a minute surface magnetic disorder significantly reduces the critical temperature ($T_\mathrm{c}$) and broadens the superconducting transition as the film thickness and device size decrease. We measure electron dephasing rates via magnetoresistance from $T_\mathrm{c}$ to $\sim 4T_\mathrm{c}$ in various-thickness TiN films. Electron dephasing occurs on the picosecond timescale and is nearly independent of temperature, differing from the expected inelastic scattering due to the electron-phonon and electron-electron interactions near $T_\mathrm{c}$, which occur over a nanosecond timescale. We propose spin-flip scattering as a possible additional phase-breaking mechanism. The significant increase in the dephasing rate for the thinnest film indicates that magnetic disorder resides near the surface of naturally oxidized films. Our research suggests that magnetic disorder may be a significant contributor to RF dissipation in superconducting devices based on TiN.
\end{abstract}

\maketitle

Recent developments in superconducting quantum systems have established them as a key platform for scalable and fault-tolerant quantum computing~\cite{Arute2019}. However, decoherence remains a major obstacle to progress in quantum information processing~\cite{Krantz2019}and other applications such as particle detection and quantum sensing~\cite{Zmuidzinas2012}. To address macroscopic decoherence in superconducting circuits, it is essential to understand material behavior at the microscale, focusing on defects and interfaces~\cite{Siddiqi2021}. Defects can introduce two-level systems (TLSs)~\cite{Müller_2019, arabi2024, huang2024,Bafia2024} or non-equilibrium quasiparticles~\cite{Siddiqi2021}, affecting charge noise~\cite{deGraaf2018} and device quality factors~\cite{Jayaraman2024}. One hypothesis proposes that TLS formation is the result of fluctuations in the superconducting order parameter caused by magnetic disorder~\cite{deGraaf2020}.
Magnetic disorder is known to significantly affect superconductivity by disrupting time-reversal symmetry, which is essential for superconducting pairing~\cite{AG_1961}. Experimental evidence indicates that magnetic disorder spontaneously forms in oxidized surface layers of thin films, breaking Cooper pairs and suppressing superconducting properties as the film thickness decreases ~\cite{Rogachev2006,Yang2020}. This can be observed through gap smearing in the density of states~\cite{Proslier08, Kuzmiak2022}, subgap peaks reminiscent of Yu-Shiba-Rusinov states~\cite{Tamir2022}, excess flux noise in superconducting resonators~\cite{Sendelbach08, Kumar2016} and short-range magnetic correlations in amorphous oxides~\cite{krasnikova2025}.

Epitaxial TiN films are gaining interest for quantum circuits due to their enhanced quantum coherence properties compared to alternatives ~\cite{Chang2013, Makise2015, Richardson2020, Faley2021, Gao2022, Amin2022, Deng2023, Wu2024, Bal2024}. Recent studies have revealed exceptional electronic parameters in these films, including a long electron mean free path limited by surface scattering and a critical temperature close to the bulk value~\cite{Saveskul}. These findings contrast with previous research on strongly disordered TiN films~\cite{Leduc2010, Sacepe2010, Driessen2012, Bastiaans2021}. Meanwhile, a detailed analysis of the thickness-dependent superconducting properties of epitaxial TiN films reveals moderate surface magnetic disorder, which results in a significant decrease in $T_\mathrm{c}$ and an additional broadening of the resistive transition~\cite{Saveskul, Baeva2022_p2}. This study aims to explore another aspect of electron transport in TiN films, the phase coherence of electrons at low temperatures. We analyze quantum corrections to conductivity above $T_\mathrm{c}$  and extract phase-breaking scattering rates $\tau_\mathrm{\phi}^{-1}$ in TiN films of varying thickness. Magnetotransport measurements, along with structural and chemical profiling,  highlight the possible contribution of surface magnetic disorder to electron dephasing in thin TiN films.

Epitaxial TiN films are grown on a \textit{c}-plane sapphire substrate at temperature of 800~\textdegree C using DC reactive magnetron sputtering from a 99.999~\% pure Ti target. The process occurs in an argon-nitrogen environment at a pressure of 5\,mTorr with an Ar:N$_2$ ratio of 2:8\,sccm. The deposition rate is 2.2\,nm/min, and the film thickness ($d$) is adjusted by varying the deposition time. Structural analysis confirms the material single-crystal homogeneity, in agreement with previous studies~\cite{Kinsey2014, Saveskul}. For magnetoresistance (MR) measurements, TiN films are patterned into 1000\,$\mu$m $\times$ 500\,$\mu$m Hall bar structures using photolithography and $SF_{6}$-plasma chemical etching. The devices with film thicknesses of  4, 10, 12 and 20\,nm are designated as MR1, MR2, MR3, and MR4, respectively, in Table~\ref{Table_1}.

\begin{figure*}[h]
    \includegraphics[scale=1]{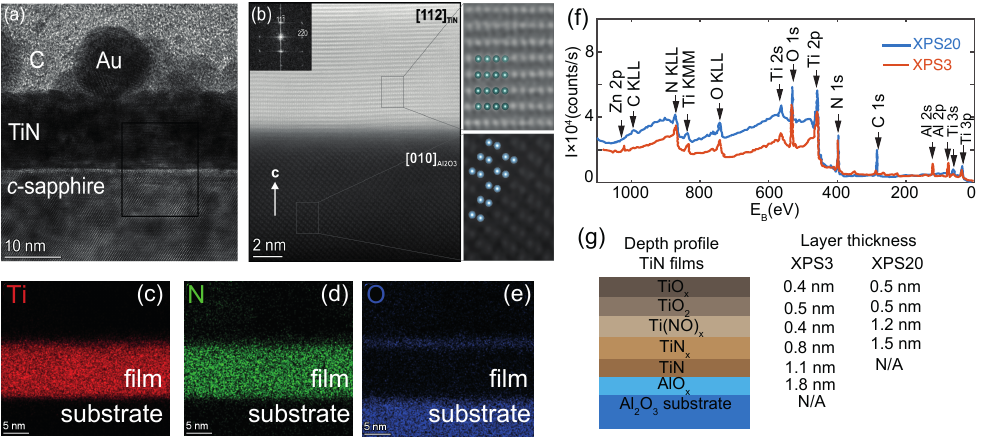}
    \caption{TiN film characterization. (a) TEM image of a 12 nm thick TiN layer on \textit{c}-sapphire. (b) High-resolution HAADF-STEM image of a TiN/Al$_2$O$_3$ interface and enlarged fragments, showing atomic packing in both structures. The projections of the structures superimposed onto the experimental images (Ti – turquoise, Al – gray). [112] Fourier transform of TiN is shown as inset in left upper corner. (c)-(e) EDX maps confirms a homogeneous distribution of Ti and N in the film and presence of thin oxidized layer on the interface. (f) Wide scan XPS spectra of 3 nm (XPS3) and 20 nm (XPS20) TiN films. (g) Chemical and phase depth profiles of XPS3 and XPS20.}
\label{fig_1}
\end{figure*}

The structure and chemical composition of the epitaxial TiN films are examined using transmission electron microscopy (TEM) and x-ray photoelectron spectroscopy (XPS) on pristine samples.  Electron diffraction patterns, conventional TEM and high-angle annular dark field scanning TEM (HAADF-STEM) images are acquired using a Titan Themis Z. Energy-dispersive X-ray spectroscopy (EDX) maps and spectra are registered in a STEM mode with an embedded Super-X detection system. The XPS analysis is performed using the electron-ion spectroscopy module based on Nanofab 25 (NT-MDT) platform. A 12-nm thick sample (TEM12) is chosen for the TEM analysis, anticipating potential epitaxial growth disruption. TEM image in Figure~\ref{fig_1}(a) illustrates 12-nm TiN film grown on the \textit{c}-sapphire substrate and coated with C/Au protective coating. Figure~\ref{fig_1}(b) shows high-resolution HAADF-STEM image of TiN/Al$_2$O$_3$ interface from the black rectangular area. Fourier-transform in the left upper corner is indexed in the [112] zone axis of the face-centered cubic structure of TiN. In the sapphire substrate, the c axis is oriented upwards, thus the TiN film is grown on its c-plane. Enlarged Fourier-filtered fragments of the HAADF-STEM image clearly visualize the atomic packing in both structures. The projections of the structures along the corresponding crystallographic directions are superimposed onto the experimental images. Thin amorphous layer between TiN and Al$_2$O$_3$ corresponds to the oxidized layer, according to the EDX elemental maps in Figure~\ref{fig_1}(c-e).  XPS studies are conducted on TiN films with 3 nm (XPS3) and 20 nm (XPS20) thicknesses. Figure~\ref{fig_1}(e) shows typical XPS spectra, revealing the presence of different phases in TiN films: TiO$_{2}$, TiO$_{x}$, Ti(NO)$_{x}$, TiN$_{x}$, TiN. Quantitative analysis determines layer composition and thickness using methods described in Ref.~\cite{Lubenchenko2018}, including background subtraction, spectral line modeling, peak decomposition, and a cascade-inhomogeneous target model. The results show an oxidized surface layer of titanium oxide and oxynitride, and the absence of any magnetic elements (Cr, Ni, Fe, etc.) due to the absence of corresponding peaks. Note that the XPS method determines the relative atomic concentration with  0.1\% accuracy.

\begin{figure}[h!]
\centering
\includegraphics[scale=1]{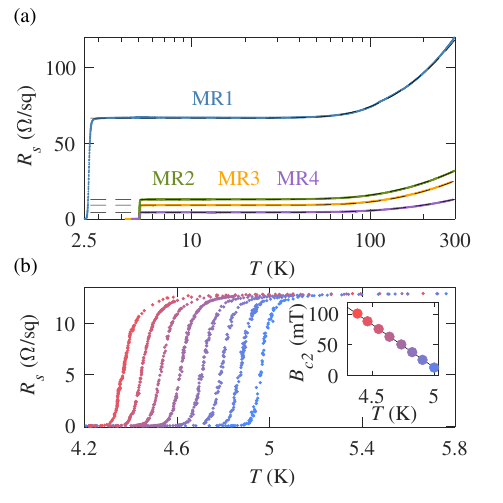}
    \caption{\label{fig_2} 
     (a) $R_\mathrm{s}(T)$ dependencies over a wide $T$ range on semi-log scale. Experimental data shown by symbols, the Bloch-Gr\"{u}neisen fits by dashed lines. (b) Main: $T$-dependencies of $R_\mathrm{s}$ on magnetic field for sample MR2. Inset: $T$-dependence of second critical magnetic field, $B_\mathrm{c2}$.}
\end{figure}

The transport properties of TiN devices are examined in a wide temperature range, as shown in Figure~\ref{fig_2}. To investigate the electron transport properties, Hall bar devices (MR1 - MR4) are mounted in a cryogenic insert immersed in a $^{4}$He dewar. The sheet resistance ($R_\mathrm{s}$) is measured in a four-probe configuration using Lake Shore 370 AC Resistance Bridge with bias current $<1$\,$\mu$A. Temperature is monitored using a calibrated diode thermometer and  Lake Shore 218 Temperature monitor. Figure~\ref{fig_2}(a) displays the $R_\mathrm{s}(T)$ dependencies for all TiN samples on a semilog scale. At higher temperatures, the data exhibits metallic behavior. At lower temperatures, $R_\mathrm{s}$ saturates at a residual resistance below 50\,K and drops to zero below $T_\mathrm{c}$, which is defined as the temperature where resistance halves compared to $R_\mathrm{s}^{10K}$, the normal-state resistance at 10\,K. We also analyze the suppression of the resistive transition in TiN samples by applying perpendicular magnetic field. Figure~\ref{fig_2}(b) shows the $R_\mathrm{s}(T)$ curves for MR2 sample. From the linear fit of the $B_\mathrm{c2}(T)$ data (inset in Figure~\ref{fig_2}(b)), we determine the zero-temperature value of second critical magnetic field $B_\mathrm{c2}(0)$, the Ginzburg-Landau coherence length $\xi_\mathrm{GL}^{2} = \Phi_0/(2\pi B_{c2}(0))$, and the diffusion coefficient $D = 4k_\mathrm{B}T_\mathrm{c}/(\pi e B_\mathrm{c2}(0))$. Here $\Phi_0$ is the magnetic flux quantum, $k_\mathrm{B}$ is the Boltzmann constant. The calculated $\xi_\mathrm{GL}$ ranges from 23 to 36\,nm. Previous studies of similar TiN films~\cite{Saveskul} found that diffusive surface scattering dominates in films thinner than 20 nm with the mean free path is about 9-16 nm for MR1-MR4, respectively. These values are approximately twice the estimate of the transport relaxation time ($\tau_\mathrm{tr}$) and the diffusion coefficient ($D$): $l_\mathrm{tr} = \sqrt{3D\tau_{tr}}$ ( $\sim 2 - 7$\,nm for MR1-MR4). For MR1, MR2, and MR4, $\tau_\mathrm{tr} = 1/(\rho^{10K}\omega_p^2\varepsilon_0)$, where $\rho^{10K} = R_\mathrm{s}^{10K}d$ and $\omega_p$ is plasma frequency. For MR3, $\tau_\mathrm{tr} = 3D/v_\mathrm{F}$, with the Fermi velocity $v_\mathrm{F} \approx 4 \times 10^5$\,m/s~\cite{Saveskul}. The values of $T_\mathrm{c}$, $B_\mathrm{c2}(0)$, $D$, $\omega_p$, $\tau_\mathrm{tr}$, and $l_\mathrm{tr}$ are summarized in Table~\ref{Table_1}. The discrepancy in the mean free path estimates arises from considering a non-conductive oxide layer in the first analysis but not separating it from TiN film properties in the second.

\begin{table*}
\caption{\label{Table_1} Relevant parameters for the epitaxial TiN samples}
\begin{ruledtabular}
\begin{tabular}{cccccccccc}
 \textnumero & $d$ (nm)  & $T_\mathrm{c}$ (K) & $R_\mathrm{s}^{10K}$ ($\Omega$/sq) & $B_\mathrm{c2}(0)$ (T) & $D$ (cm$^{2}$/s)& $\omega_p$ (eV)& $\tau_\mathrm{tr}$ (fs) & $l_\mathrm{tr}$ (nm) &$\tau_\mathrm{s}$ (ps) \\
\hline
MR1 &  4 & 2.65 & 66.5 & 0.46 & 6.3 & 7.01 & 3.8 & 2.7 & 2.3\\
MR2 & 10 & 5.10 & 12.9 & 0.72 & 7.8  & 6.85 & 8.1 & 4.4& 12.2\\
MR3 &  12 & 5.14 & 9.1  & 0.56 & 10.08 & -  &18.9 \footnotemark[1]& 7.6 & 13.2 \\
MR4 & 20 & 5.09 & 4.4  & 0.37 & 15.3 & 7.02 & 11.3 & 7.2 & 11.9\\
MR5\footnotemark[2] &  12 & 4.6 & 9.4  & 0.43 & 12.79  & -  & 24.0 & 9.6& 6.2\\
\end{tabular}
\end{ruledtabular}
\footnotetext[1]{$\tau_\mathrm{tr} = 3D/v_\mathrm{F}$}
\footnotetext[2]{TiN sample with 1-nm thick Cr layer.}
\end{table*}

\begin{figure}[h!]
    \centering
    \includegraphics[scale=1]{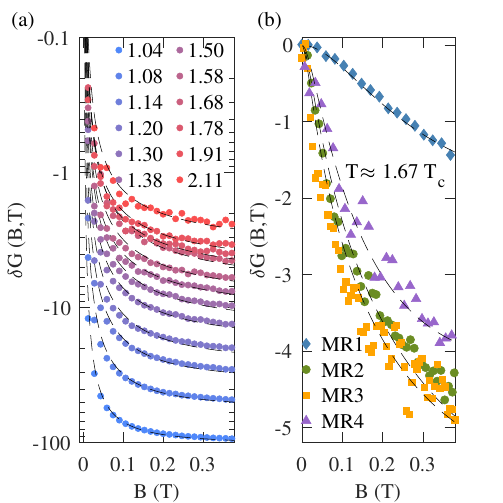}
    \caption{\label{fig_3} 
    The dimensionless magnetoconductance, $\delta G(B,T)$, is plotted versus magnetic field for a representative sample (MR3) at various bath temperatures (a) and compared for four samples (MR1 - MR4) at a fixed $T/T_\mathrm{c} \approx 1.7$ (b). Experimental data are represented by symbols, and black lines show the best fits using Eq.\eqref{eq:fit_MC_1}.}
\end{figure}

To study low-temperature electron dephasing in epitaxial TiN films, we measure magnetoresistance, $R_\mathrm{s}(B,T)$, at different bath temperatures. Figure~\ref{fig_3}(a) shows data for sample MR3, represented by dimensionless magnetoconductance $\delta G(B,T) = (2\pi^2 \hbar)/e^2\left[R_\mathrm{s}(B,T)^{-1} - R_\mathrm{s}(0,T)^{-1}\right]$, where $R_\mathrm{s}(0,T)$ is the magnetoresistance in a zero magnetic field. We fit the experimental $\delta G(B,T)$ by the relative magnetoconductance $\delta G^\mathrm{QC}= G^\mathrm{QC}(B,T)- G^\mathrm{QC}(0,T)$, using the quantum corrections to magnetoconductance~\cite{Glatz2011, Rosenbaum1985, LS_1985} (see details in Appendix). Taking into account the experimentally determined values of $T_\mathrm{c}$, $D$, $B_\mathrm{c2}(0)$, $\tau$, and $\tau_\mathrm{s}$ (see estimates below), we fit the experimental data for $\delta G(B,T)$ considering the electron phase-breaking time $\tau_\mathrm{\phi}$ as a fitting parameter. Details of the fitting procedure are given in Appendix.  

\begin{figure}[h!]
   \includegraphics[scale=1]{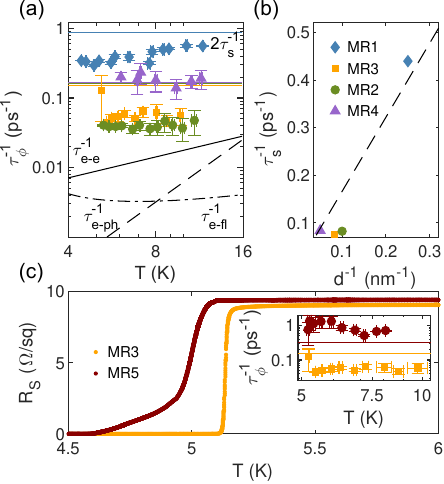}
    \caption{\label{fig_4} Phase-Breaking Rates and Scattering Processes in Epitaxial TiN Films. (a) Electron phase-breaking rate $\tau_\phi^{-1}$ for all TiN samples. MR1 data are blue diamonds, MR2 orange squares, MR3 green circles, MR4 purple triangles. Data are plotted in log-log scaled. Solid colored lines show $2\tau_\mathrm{s}^{-1}$ estimates from the AG model using experimental $T_\mathrm{c}$ values. Black lines indicate contributions from the e-ph (dashed), e-e (solid), and e-fl (dot-dashed) scattering for the thinnest sample (MR1). Error bars for $\tau_\mathrm{\phi}^{-1}$ are obtained from temperature measurement uncertainty, estimated from resistance fluctuations at $B = 0$. (b) Estimated spin-flip scattering rate $\tau_\mathrm{s}^{-1}$ vs. inverse thickness $d^{-1}$. Dashed line shows a fit to the data, indicating that the dominant spin-flip scattering originates from near-surface magnetic disorder. (c) Main: $R_\mathrm{s}$ dependencies for MR3 and MR5 samples (MR5 has a 1-nm Cr layer on TiN). Inset: Experimental $\tau_\mathrm{\phi}^{-1}$ values for MR5, compared to initial MR3 data and AG model predictions.}
\end{figure}

The main result of this study is shown in Figure~\ref{fig_4}(a), which illustrates the phase-breaking rate, $\tau_\mathrm{\phi}^{-1}$, for epitaxial TiN films of varying thicknesses. As shown in Figures~\ref{fig_3}(b) and ~\ref{fig_4}(a), the experimental data depend on the film thickness. The smaller $d$, the lower $\delta G(B,T)$ and, therefore, the higher values of $\tau_\mathrm{\phi}^{-1}$ are observed. The values of $\tau_\mathrm{\phi}$ range from 2\,ps (MR1) to 10 - 30\,ps (MR2 - MR4), without a pronounced temperature dependence.

In the following, we discuss potential factors that may contribute to findings in epitaxial TiN films. Electron dephasing above $T_\mathrm{c}$ is attributed to inelastic and magnetic scattering. Inelastic scattering includes electron-electron (e-e), electron-fluctuation (e-fl), and electron-phonon (e-ph) scattering. The combined effect is described by $\tau_\mathrm{\phi}^{-1} = \tau_\mathrm{e-e}^{-1} + \tau_\mathrm{e-fl}^{-1} + \tau_\mathrm{e-ph}^{-1} + 2\tau_\mathrm{s}^{-1}$. Previous studies of disordered TiN films with high sheet resistance~\cite{Baturina_2012, Yadav2023} ($R_\mathrm{s}\sim 1\,k\Omega$) show that the phase-breaking rate is limited by e-e scattering~\cite{AA1985}: $\tau_\mathrm{e-e}^{-1} = (\pi g k_\mathrm{B} T)/\hbar\ln\left(1/(2\pi g)\right)$, where $g = e^2R_\mathrm{s}/(2\pi^2\hbar)$. In epitaxial TiN, $\tau_\mathrm{e-e}^{-1}(T)$ exceeds 60\,ps at $T < 10$ K (sample MR1), and this contribution accounts for approximately 5\% of the observed $\tau_\mathrm{\phi}^{-1}$ (Figure~\ref{fig_4}(a)).
Electron scattering due to superconducting fluctuations is expected to be have a pronounced upturn in $\tau_\mathrm{\phi}^{-1}$ as $T$ approaches $T_\mathrm{c}$. However, as shown in Figure~\ref{fig_4}(a), the lack of a characteristic temperature dependence suggest its negligible contribution to the observed results. The low temperature e-ph scattering rate on acoustic phonons follows $\tau_\mathrm{e-ph}^{-1} = 7\pi\zeta(3)\lambda_\mathrm{3D}k_\mathrm{B}T^{3}/(2\hbar\theta_\mathrm{D}^{2})$~\cite{Pethick1979}. Using the Debye temperature $\theta_\mathrm{D} = 475 \pm 25$\,K~\cite{Saveskul} and the e-ph coupling constant $\lambda_\mathrm{3D} \approx 0.73$~\cite{Mei2013}, the estimates of $\tau_\mathrm{e-ph}$ ranges from 0.1 to 10\,ns at 10 - 3\,K (Figure~\ref{fig_4}(a)). These results are consistent with photoresponse measurements on disordered TiN ($\tau_\mathrm{e-ph} \sim$ 5.5 - 88\,ns at 4.2 - 1.7\,K~\cite{Kardakova2013}) and noise measurements on epitaxial TiN ($\sim$ 5.0 - 30\,ns at 4.0 - 2\,K~\cite{Baeva2022_p1}). Thus, electron-phonon scattering, which contributes less than 10\% to $\tau_\phi^{-1}$, has a negligible effect on electron dephasing in TiN films.

We consider another important mechanism of electron dephasing that is associated with scattering by magnetic disorder. This process results in an increase in $\tau_\mathrm{\phi}^{-1}$ and a temperature-independent behavior of $\tau_\mathrm{\phi}^{-1}$ at low temperatures~\cite{LinBird_2002}. As previously demonstrated in epitaxial TiN films~\cite{Saveskul}, a moderate level of magnetic disorder significantly suppresses $T_\mathrm{c}$ compared to other typical mechanisms such as Coulomb interactions, reduced carrier density, and BCS coupling. Using Abrikosov-Gor'kov theory \cite{Abrikosov2017}, we estimate the spin-flip scattering rate $\tau_\mathrm{s}^{-1}$, assuming that $T_\mathrm{c}$ is controlled by magnetic disorder: $\ln\left(T_\mathrm{c}^0/T_\mathrm{c}\right) = \Psi\left(1/2 + \hbar/(2\pi k_\mathrm{B} T_\mathrm{c} \tau_\mathrm{s})\right) - \Psi\left(1/2\right)$, where $T_\mathrm{c}^0$ is the critical temperature in the absence of magnetic disorder. For epitaxial TiN films with a highest observed transition temperature of 5.6\,K~\cite{Saveskul}, the estimated values of $\tau_\mathrm{s}$ for MR1-MR4 samples are 2.3\,ps, 12.2\,ps, 13.2\,ps and 11.9\,ps, respectively. Figure~\ref{fig_4}(b) shows the estimated spin-flip scattering rate $\tau_\mathrm{s}^{-1}$ as a function of an inverse thickness $d^{-1}$. Note that the variation of $\tau_\mathrm{s}^{-1}$ with thickness suggests the presence of near-surface magnetic disorder. Using a simple model~\cite{Saveskul}, we relate $\tau_\mathrm{s}^{-1}$ to the magnetic scatterer density $N_\mathrm{M} = a/(v_\mathrm{F} \tau_\mathrm{s})$, where $N_\mathrm{M}$ includes both bulk $N_\mathrm{b}$ and surface $N_\mathrm{s}$ contributions: $N_\mathrm{M} = N_\mathrm{b} + N_\mathrm{s} a/d$. Here $a = 0.4$\,nm is the TiN lattice constant. The dashed line in Figure~\ref{fig_4}(b) shows the model fit with $N_\mathrm{b} = 1 \times 10^{-5}$, $N_\mathrm{s} = 3.9 \times 10^{-3}$, and $N_\mathrm{M}$ < 0.03\%. These values are below XPS detection limits. In Figure~\ref{fig_4}(a), the experimental data for $\tau_\mathrm{\phi}^{-1}(T)$ are compared with $2\tau_\mathrm{s}^{-1}$, which provides a closer estimate of the experimental data compared to the inelastic contribution. The value of $N_\mathrm{s}$ also provides an estimate of the surface density of magnetic defects, approximately $N_\mathrm{s}a^2\approx 2\times10^{12}$\,cm$^{-2}$, comparable to the typical surface density of magnetic moments ($n \sim 5\times10^{13}$\,cm$^{-2}$) in Al, Nb and NbN superconductors~\cite{Koch2007, Sendelbach08, Kumar2016, deGraaf2018}. 

The additional deposition of a 1-nm paramagnetic Cr layer on top of MR3 sample showed a shifted $R_s(T)$ curve to lower temperatures, a broader resistive transition, and a significant increase in $\tau_\mathrm{\phi}^{-1}$ as well (Figure~\ref{fig_4}(c) for MR5 sample). Although the estimated values of $\tau_\mathrm{s}^{-1}$ are closer to experimental results than those due to inelastic scattering, they are still not accurate enough to fully explain the observed phenomena. This may be due to the use of a too simple model for estimating $\tau_\mathrm{s}^{-1}$, and further research is needed.

The origin of magnetic disorder in nominally non-magnetic Ti-based materials is an active research area. Stoichiometric TiN, typically considered as a Pauli paramagnet~\cite{Allmaier2009}, may exhibit magnetic properties due to point defects like nitrogen vacancies~\cite{Gupta2019}. In Ti-based oxides, magnetic disorder may originate from unpaired electrons associated with oxygen vacancies~\cite{Venkatesan2004}. For instance, oxygen vacancies in TiO$_2$ result in magnetic $Ti^{+3}$ and $Ti^{+2}$ ions~\cite{Yoon2006}, which are also part of non-stoichiometric oxides on the TiN film surface (see the supplemental file in Ref.~\cite{Saveskul}). Nitrogen-doped, non-stoichiometric TiO$_2$ may exhibit magnetic behavior due to reduced band gap and overlapping oxygen-vacancy states with the empty conduction band~\cite{Drera2010, Zhou2009}.

To minimize the impact of surface magnetic disorder on device performance, it is essential to implement passivation methods, which involve in situ surface treatments to prevent oxide formation. For example, passivated Nb and Cu films exhibit longer dephasing times, with a power-law dependence on temperature ($T^{-2.5}$ for Nb~\cite{Lomakin2022} and $T^{-3}$ for Cu \cite{Vranken1988}), compared to unpassivated films, where $\tau_\mathrm{\phi}^{-1}(T)$ saturates at low temperatures. Other examples include depositing an AlScN layer on epitaxial TiN to decouple the effects of oxidation and electron confinement~\cite{Shah2022}. Depositing a silicon layer on NbN produces ultra-thin films with high critical temperatures~\cite{Lomakin2023}. Metal capping layers, such as Ta and TiN, significantly improve the coherence times of Nb-based qubits~\cite{Bal2024}.

In summary, we investigate the quantum corrections to conductivity in the temperature range from $T_\mathrm{c}$ to $\sim 4T_\mathrm{c}$, revealing electron phase-breaking rates almost independent of temperature in epitaxial TiN films. Our analysis shows that surface magnetic disorder can be a limiting factor in the electron dephasing in thin films.


\begin{acknowledgments}
The authors would like to express their sincere gratitude to V.S. Khrapai for his constructive comments and valuable insights. Additionally, the authors acknowledge the significant contributions of A. Boltasseva and V.M. Shalaev in providing thin film samples and their assistance in the preparation of the manuscript. This work was funded by the RSF grant 24-72-10105 (the transport measurements) and the Basic Research Program at the HSE University (the analysis of magnetoresistance data). The TEM study was done on the facilities of AICF, Skoltech. The fabrication of samples was supported by the "Priority 2030 Program" of the National University of Science and Technology MISIS.
\end{acknowledgments}

\section*{Data Availability Statement}

The data that support the findings of this study are available from the corresponding author upon reasonable request.

\appendix
\section{Magnetoresistance in thin films}
For two-dimensional superconducting films, $\delta G^\mathrm{QC}$ is expected to be the sum of four terms: the Aslamazov-Larkin (AL) term, the density of states (DOS) contribution term, the Maki-Thompson (MT) term, and the weak localization (WL).

\begin{align}
	\label{eq:fit_MC_1}
    \scalebox{0.9}{$
               G^\mathrm{QC}(B,T)  = \underbrace{ \frac{\pi^2\epsilon}{4h^2}\left[\psi\left(\frac{1}{2}+\frac{\epsilon}{2h}\right)-\psi\left(1+\frac{\epsilon}{2h}\right)+\frac{h}{\epsilon}\right]}_{\rm{AL}}$ \nonumber}\\
\scalebox{0.9}{$\underbrace{-\frac{28\zeta\left(3\right)}{\pi^2} \left[\ln \left(\frac{1}{2h}\right) -\psi\left(\frac{1}{2}+\frac{\epsilon}{2h}\right)\right]}_{\rm{DOS}}$ \nonumber}\\
\scalebox{0.9}{$-\underbrace{\beta_{MT}(T,\tau_\phi) \left[\psi\left(\frac{1}{2}+\frac{B_\phi}{B}\right)-\psi\left(\frac{1}{2}+\frac{B_\phi}{B}\frac{\epsilon}{\gamma_{\phi}}\right)\right]}_{\rm{MT}}$ \nonumber}\\
\scalebox{0.9}{$\underbrace{+\frac{3}{2}\psi\left(\frac{1}{2}+\frac{B_2}{B}\right)-\psi\left(\frac{1}{2}+\frac{B_1}{B}\right)-\frac{1}{2}\psi\left(\frac{1}{2}+\frac{B_\mathrm{\phi}}{B}\right)}_{\rm{WL}}.$\nonumber}\\
\end{align}
The given expressions are applied for two-dimensional (2D) systems, where the film thickness, $d$, is smaller than the characteristic length scales: the thermal coherence length $L_\mathrm{T} = \sqrt{2\pi\hbar D/(k_\mathrm{B}T)}$, the phase-breaking length $L_\mathrm{\phi} = \sqrt{D\tau_\mathrm{\phi}}$ and the superconducting coherence length $\xi_\mathrm{GL}$ ($d < L_{T}, L_\phi, \xi_{GL}$). Here $\psi(x)$ is the Digamma function, $\epsilon = \ln(T/T_\mathrm{c})$ and $h = 0.69 B/B_\mathrm{c2}(0)$ are the reduced temperature and magnetic field, respectively, $\gamma_\mathrm{\phi} = \pi\hbar/(8k_\mathrm{B}T\tau_\mathrm{\phi})$ is the phase-breaking parameter. The characteristic fields are defined as $B_1 = B_\mathrm{0} + B_\mathrm{so} + B_\mathrm{s}$,  $B_2 = B_\mathrm{\phi} + \frac{4}{3}(B_\mathrm{so} - B_\mathrm{s})$, $B_{0} = \hbar/(4eD\tau)$, $B_\mathrm{so} = \hbar/(4eD\tau_\mathrm{so})$, $B_\mathrm{s} = \hbar/(4eD\tau_\mathrm{s})$, and $B_\mathrm{\phi} = \hbar/(4eD\tau_\mathrm{\phi})$, where $\tau$, $\tau_\mathrm{so}$, $\tau_\mathrm{s}$, and $\tau_\mathrm{\phi}$ are the corresponding times for the elastic, the spin-orbit, the spin-flip and the phase-breaking scattering, respectively. The coefficient in the MT term $\beta_{MT}(T,\tau_\mathrm{\phi})$ is given in Ref.~\cite{LS_1985}. Note that the magnetoresistance at high temperatures ($\epsilon > 1, T \gtrsim 3T_\mathrm{c}$) is mainly determined by the MT and WL terms in Eq.\eqref{eq:fit_MC_1}, while the contribution of the AL and DOS terms in Eq.\eqref{eq:fit_MC_1} is significant at low temperatures ($\epsilon \ll 1$).

In further analysis, the spin-orbit scattering time is estimated as $\tau_\mathrm{so} = \tau(\alpha Z)^{-4}$, where $\alpha$ is the fine structure constant, and $Z$ is the effective atomic number of material ($Z_\mathrm{TiN} = 14.5$). The estimated values are 48\,ps, 88\,ps, 120\,ps, and 143\,ps for MR1, MR2, MR3, and MR4, respectively. The phase-breaking length, defined as  $L_\mathrm{\phi} = \sqrt{D\tau_\mathrm{\phi}} \approx 30 - 150$\,nm, exceeds $d$ of the studied samples for the considered $T$ range. This fact supports the validity of using Eq. \eqref{eq:fit_MC_1} for the 2D case. 

\nocite{*}
\bibliography{m_bibliography}

\end{document}